\documentclass[aps,amssymb,amsmath,superscriptaddress,showkeys,prl, twocolumn]{revtex4-1}

\usepackage[utf8x]{inputenc}
\usepackage{graphicx}
\usepackage{epsf}
\usepackage{units}
\usepackage{ulem}

\usepackage{hyperref}
\usepackage{amssymb,amsmath}
\usepackage{graphicx}
\usepackage{braket}
\usepackage{color}
\usepackage{xcolor}

\usepackage{graphicx}
\usepackage{units}
\usepackage[mediumspace,mediumqspace,squaren]{SIunits}

\newcommand{\PP}[1]{probe particle }

\graphicspath{{IMG/}}

\begin{document}

\title{Principles and simulations of high-resolution STM imaging with flexible tip apex.}

\author{Ondrej Krej\v{c}\'{i}}
\email[corresponding author: ]{krejcio@fzu.cz}
\affiliation{Institute of Physics of the Czech Academy of Sciences, v.v.i.,  Cukrovarnick\' a 10, 162 00 Prague, Czech Republic}
\affiliation{Charles University in Prague, Faculty of Mathematics and Physics, Department of  Surface and Plasma Science, V Hole$\check s$ovi$\check c$k\'ach 2, 180 00, Prague, Czech Republic}

\author{Prokop Hapala}
\affiliation{Institute of Physics of the Czech Academy of Sciences, v.v.i.,  Cukrovarnick\' a 10, 162 00 Prague, Czech Republic}

\author{Martin Ondr\'{a}\v{c}ek}
\affiliation{Institute of Physics of the Czech Academy of Sciences, v.v.i.,  Cukrovarnick\' a 10, 162 00 Prague, Czech Republic}

\author{Pavel Jel\'{i}nek}
\affiliation{Institute of Physics of the Czech Academy of Sciences, v.v.i.,  Cukrovarnick\' a 10, 162 00 Prague, Czech Republic}
\affiliation{Donostia International Physics Center (DIPC), Paseo Manuel Lardizabal 4, E-20018 San Sebastian, Spain}


\begin{abstract}

We present a~robust but still efficient simulation approach for high-resolution scanning tunneling  microscopy with a~flexible tip apex showing sharp submolecular features. The approach takes into account the electronic structure of sample and tip and relaxation of the tip apex.
We validate our model by achieving good agreement with various experimental images which allows us to explain the origin of several observed features. Namely, we have found that high-resolution STM mechanism consists of the standard STM imaging, convolving electronic states of the sample and the tip apex orbital structure, with the contrast heavily distorted by the relaxation of the flexible apex caused by interaction with the substrate.

\end{abstract}

\maketitle
Both scanning tunneling microscopy (STM) \cite{Temirov_NJP08} and atomic force microscopy (AFM) \cite{Gross_Science09} demonstrated capability to achieve the submolecular resolution with properly functionalized tip apex \cite{Bartels_PRL98, GrossAPL2013}. The unprecedented spatial resolution das advanced a~characterization of atomic clusters \cite{Emmrich_Science15}, single molecules \cite{Gross2012, AlbrechtPRL2015, Hapala_NC_2016}, their assemblies \cite{Hamalainen2014, Kawai_ACSNano2013} \iffalse,Kawai_ACSNano2015} \fi and mutual interactions \cite{Sun_2011, Corso_PRL2015} as well as the understanding of on-surface chemical reactions \cite{Gross_NatChem_2010,PavlicekNatChem2015} including an~identification of intermediate and final products \cite{DeOteyza2013, SchulerJACS2015}. 

The origin of the submolecular AFM contrast is well estabilished \cite{Moll_NJP10, Hapala_PRB_2014, Hamalainen2014, Guo_JPCC2015}. In general, the sharp submolecular contrast results from a~lateral bending of a~flexible tip apex (in our approach represented by a~probe particle \cite{Hapala_PRB_2014} - PP). This bending is caused  by a~lateral force acting on the tip apex, which results from an~interplay between repulsive Pauli, attractive van der Waals and electrostatic forces \cite{Hapala_NC_2016}. Sharp edges representing apparent bonds in AFM images \cite{Gross2012,Pavlicek_PSSB2013} are the consequence of a discontinuity in the lateral bending of the flexible apex above ridgelines of the potential energy landscape, which are typically located above atoms or bonds \cite{Hapala_PRB_2014,Hamalainen2014,Boneschanscher2014}.
 
In principle, the high resolution STM (HR-STM) imaging represents an~experimentally a~less demanding way to achieve submolecular contrast than AFM. Furthermore, it provides information about the electronic structure, in addition to the physical structure of the inspected molecules Thus, information provided by STM is in principle superior to AFM. However, a~detailed understanding of the HR-STM imaging mechanism is still missing \cite{Martinez_PRL2012, Weiss2010a, Hapala_PRB_2014}, which impedes its wider proliferation.

Previously, we demonstrated  \cite{Hapala_PRB_2014, Hapala_PRL_2014} that the relaxation of the flexible PP attached to the tip can partially explain the submolecular contrast observed not only in AFM, but also in STM and inelastic electron tunneling spectroscopy (IETS) \cite{Chiang_Science2014} images . However, the original STM model \cite{Hapala_PRB_2014} neglects completely the electronic structure in the description of the tunneling process between tip and sample. The fact that such crude tunneling model was able to reproduce to some extend  the sharp features visible in HR-STM experiments further emphasize the importance of accounting for the tip apex relaxation in the close distance regime. On the other hand, numerous HR-STM experiments \cite{Temirov_NJP08, Weiss2010a, Hapala_Book_2015} indicate that the submolecular contrast depends very much on various experimental details - such as the bias voltage, the substrate or the microscopic structure of STM tip apex. Thus, it is evident that the electronic structure has to be included in the correct description of the HR-STM imaging.

Traditional STM simulation methods are based on either non-perturbative approach \cite{Cerda_PRB97,Mingo_PRB96, Blanco2006}, or perturbative approach (e.g.\cite{Hofer2003}). The later is only valid when the tip and sample remain out of the tip-sample physical contact \cite{BlancoGonzalez2004}. It frequently uses the Bardeen approach \cite{Bardeen1961} and subsequent approximations derived by Chen \cite{Chen_paper,Chen_book,Palotas_Chen} or Tersoff and Hamann \cite{TersoffHamann1985} (TH). Importantly, the STM methods were  devised for the surfaces of solid states materials with a~rigid tip apex. Thus they do not take into account any tip apex relaxation, which is fundamental for the understanding of the submolecular contrast with functionalized tips. 

In this work, we present an~efficient STM model, which takes into account both the PP relaxation as well as electronic wave functions of tip and sample. We will show that the new model (hereafter referred to as PP-STM) \cite{Github_stm} is able to explain experimentally observed features, which could not be properly reproduced with either the original simple model \cite{Hapala_PRB_2014} or the traditional STM methods. 

High resolution AFM/STM images with functionalized tips are typically acquired at very close distances where repulsive Pauli forces  dominates. Therefore, the influence of the tip proximity can substantially affect the tunneling barrier \cite{BlancoGonzalez2004}. Nevertheless, it has been shown \cite{Corso_PRL2015} that the tunneling barrier is preserved even in the repulsive regime, due to the presence of a low-reactive functionalized tip apex, such as CO or Xe. Consequently the perturbative approach, describing tunneling processes, still remains valid.

Thus in our model, we adopt the Bardeen \cite{Bardeen1961} based approach to express the tunneling current $I$ between different eigenstates of the sample $S$ and the tip $T$ (in atomic units):
\begin{align}
 I = 4\pi \int_{0}^{V} \sum_{T} \sum_{S} \rho_{T}(E_F-V+\nu)\rho_{S}(E_F+\nu) |M_{TS}|^2 d\nu,
\label{eq_low}
\end{align}
where $\rho_T,\rho_S$ means densities of states (DOS) of tip and sample, respectively; $V$ represents applied bias voltage and $M_{TS}$ the tunneling matrix.  The tunneling matrix elements - $M_{TS}$ - are approximated by the so called Chen's rules \cite{Chen_paper,Chen_book}. The electronic structure of the sample is expressed in Local Combinations of Atomic Orbitals (LCAO) formalism, and the wave-function coefficients $c^{\text{LCAO}}_{S,a,\alpha}$ are obtained from the DFT calculations \cite{Review-Fireball2011,GPAW1,GPAW2,GPAW-LCAO,AIMS}. The atomic radial functions are approximated by an~exponential function with the characteristic decay length $\kappa$ determined by the work function of the sample. The tunneling matrix elements $M_{TS}$ are calculated by:
\begin{align}
M_{TS} = \sum_a 4\pi C_a \kappa^{1/2} exp(-\kappa |\vec{r}_a|) \sum_\alpha Y_{T\alpha} c^{\text{LCAO}}_{S,a,\alpha},
\label{eq_MTS}
\end{align}
where the summation goes over atoms of the sample $a$ and corresponding atomic orbitals $\alpha$ of atom $a$. $Y_{T\alpha}$ is a~rational function originating from the Chen's rules. $|\vec{r}_a|$ stands for distance between the atom $a$ and PP; $C_a$ is an~amplitude constant. Unless stated otherwise, the decay $\kappa$ and the amplitude constant $C_a$  are assumed to be the same for all the atoms of the sample. 
The sample DOS $\rho_S$ is obtained from the eigenstates taken from total energy DFT calculations of the sample. For simplicity, we consider the tip wave functions being represented by a~non-tilting atomic orbitals - $s$, $p_x$, $p_y$ and $p_z$, each of which serve as an~independent tunneling channel. Therefore, $\rho_T$ is a~parameter fitted to the experimental results. The atomic orbitals are located on the relaxing PP, whose positions are pre-calculated via PP-AFM code \cite{Hapala_PRB_2014, Hapala_PRL_2014}.
In case of weakly bound tip apex (e.g. Xe atom) two tunnelings can appear - between sample and PP and between PP and tip \cite{Hapala_PRB_2014}. Here we consider only the tunneling between the PP and the sample. The validity of this approach is supported by good agreement with experimental evidence as discussed later. However, we cannot rule out that in some cases the second tunneling can further modulate the calculated signal. More detailed description of the STM model can be found in \cite{SI}.


In what follows, we will examine our approach by comparison with experimental HR-STM images obtained above three different systems: perylenetetracarboxylic dianhydride (PTCDA) molecule adsorbed on Au(111) \cite{Kichin2013, Martinez201114} and Ag(111) \cite{Hapala_NC_2016, Rohlfing2007} surfaces and 1,5,9-trioxo-13-azatriangulene (TOAT) molecule adsorbed on Cu(111) surfaces \cite{Nadine_submitted}. Details of the total energy DFT calculations of the systems are described in \cite{SI}; the underlying methodology is described in \cite{Review-Fireball2011,GPAW1,GPAW2,GPAW-LCAO,Kresse1996,vanderbilt1990,Grimme2006,Jorgensen1988}.The experimental measurements were done in a~constant height STM mode with very low applied bias voltage or in a~constant height dI/dV mode. Since the experimental images acquired in very low bias voltage can be seen as dI/dV maps, all simulated STM images were calculated as constant height dI/dV maps at a~particular energy.

\begin{figure}
\centering
\includegraphics[width=8.5cm]{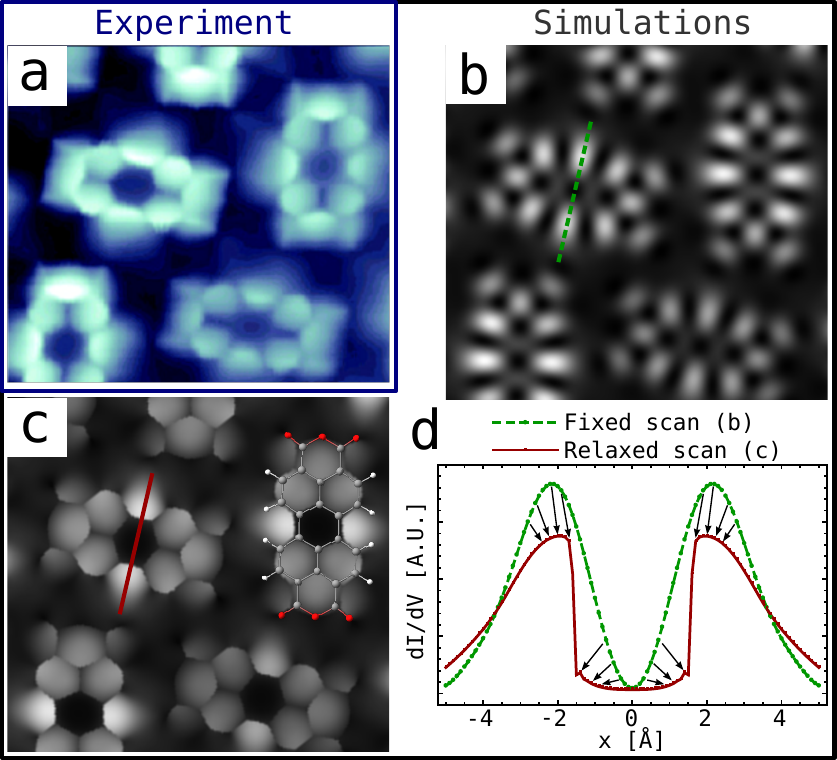}
\caption{ \textbf{Comparison between experimental and theoretical STM images calculated with rigid and flexible tip apex:} (a) Experimental constant height HR-STM dI/dV figure of PTCDA/Au(111) obtained with CO tip at $V_{bias}$~=~-1.6~V with respect to the sample \cite{Kichin2013}. (b-c) Constant height dI/dV simulations of PTCDA/Au(111) at the energy of HOMO of PTCDA obtained via our PP-STM code using $p_x$ and $p_y$ orbitals on the probe particle (PP) originally placed $3.2~$\AA~above the molecule:  with the fixed (b) and relaxed (c) PP, respectively. (d) Comparison of line profiles, taken above centers of PTCDA molecules as indicated in (b) and (c) by green dashed for fixed and red full line for relaxed PP, respectively. The arrows indicate the changes in the dI/dV signal given by the PP relaxations.}
\label{figPTCDAau}
\end{figure}

Fig.~\ref{figPTCDAau}~(a) shows the experimental dI/dV map of  PTCDA/Au(111) acquired with CO-terminated tip at the energy corresponding to the HOMO of PTCDA \cite{Kichin2013}. The molecular skeleton is rendered by the characteristic sharp edges, with a~pronounced depression of the dI/dV signal located in the central part of the molecule. To disentangle the effect of the electronic structure and the PP relaxation, we first calculated a dI/dV map at the energy of HOMO without the PP relaxation at a close tip-sample distance, see Fig.~\ref{figPTCDAau}~(b). The combination of the HOMO orbital (shown in \cite{SI} in Fig.~S2~(a)) with $p_x$ and $p_y$ orbitals on the fixed PP transforms the original  twelve lobes of the orbital into 5 stripes at each side of the molecule and 4 squares in the middle of it. It is noteworthy that unlike the TH, our approach takes into account the cancellation of an STM signal, due to interference effects \cite{Gross2011,Kolja_NGr_2014,Palotas_Chen}, that plays an important role in the formation of the STM signal. For example, the destructive interference takes place in the middle of the molecule, where the different phases of the sample and tip orbitals leads to a cancellation of the calculated signal. However, the calculated STM image with fixed PP (Fig.~\ref{figPTCDAau}~(b))  lacks  the sharp edges and overall agreement with the experimental counterpart is poor. 

In the next step, we perform STM simulations including the PP relaxation with the same tip-sample distance and energy, see Fig.~\ref{figPTCDAau}~(c). The impact of the PP relaxation is substantial and the resulting STM image agrees very well with the experimental evidence: compare Fig.~\ref{figPTCDAau}~(a) to (c). Namely, the PP relaxations distorts the smooth signal giving rise to the sharp edges above the potential ridges. The effect of the relaxation is even better pronounced on a comparison of line profiles taken above the center of the molecule, see Fig.~\ref{figPTCDAau}~(d). When the PP is located above a~central hexagon it relaxes towards its center to minimize the interaction energy (positions of the PP are shown in Fig.~S2~(d) in \cite{SI}). Therefore the signal taken above the central hexagon is almost constant.  The sharp edges are also visible on the image obtained with $s$ orbital on the relaxing PP (Fig.~S5~(a) in \cite{SI}); however, this simulation does not match well with the experimental image. Conversely, the very good agreement between the image simulated with $p_x$ and $p_y$ orbitals on the PP (Fig.~\ref{figPTCDAau}~(c)) and the experimental image obtained with a CO tip in the dI/dV mode (Fig.~\ref{figPTCDAau}~(a)) indicates, that the electronic structure of the CO tip  in this experiment can be well described with $p_x$ and $p_y$ orbitals. It is noteworthy that the PP-STM represents a very efficient method, as the calculated STM images Fig.~\ref{figPTCDAau}~(c) including about 1200 atoms, takes only 1 hour on a~standard workstation.


\begin{figure}
\centering
\includegraphics[width=8.5cm]{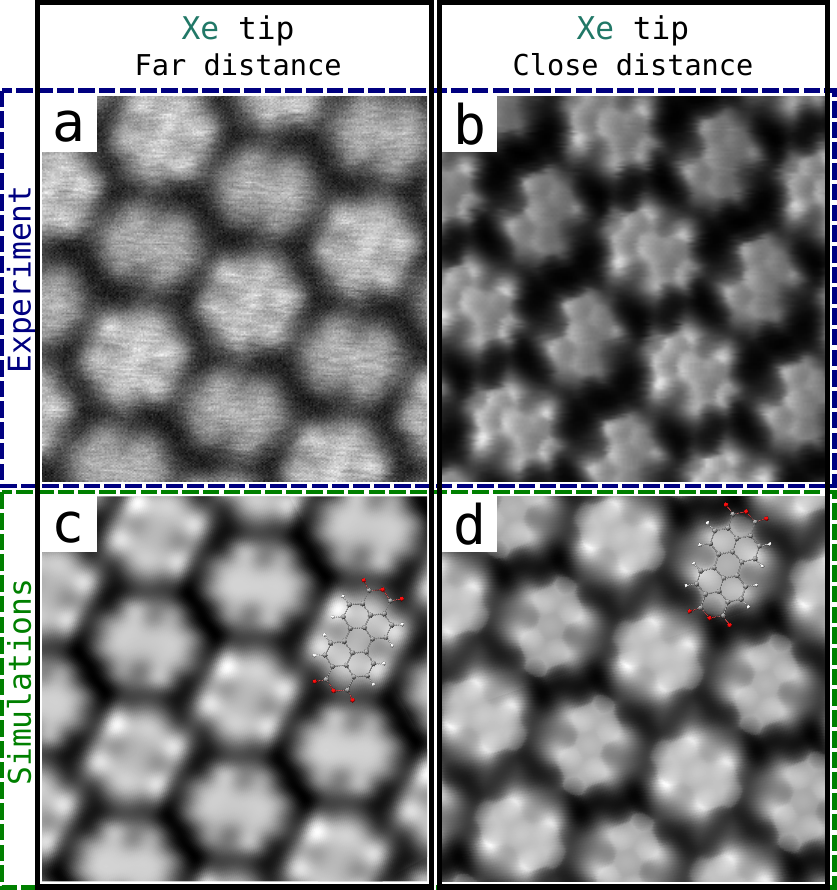}
\caption{ \textbf{Comparison of the PP-STM model with experiment - PTCDA/Ag(111) - for different heights of the tip}: (a-b) Constant height HR-STM obtained with Xe tip and $V_{bias}$~=~-2~mV in a~far (a) and close (b) tip-sample distances, respectively. (c-d) Simulated PP-STM (dI/dV) images with $s$ orbital on the PP, which is placed ~4.3~\AA~(c) and ~3.7~\AA~(d) above the molecule. The simulations energy +0.1 eV above the Fermi level, PP parameters: Q~=~+0.3~$e$ and K~=~0.2~N/m. All figures show area $39~\times~39~$\AA.}
\label{figPTCDAag}
\end{figure}


Fig.~\ref{figPTCDAag}~(a,b) displays experimental images of PTCDA molecules adsorbed on an Ag(111) surface obtained with Xe tip at low sample bias voltage ($V_{bias}$ = -2~mV \cite{Hapala_NC_2016}) in two different tip-sample distances. Fig.~\ref{figPTCDAag}~(a) was acquired in a far tip-sample distance, when tip apex relaxation is not expected. The second experimental image (Fig.~\ref{figPTCDAag}~(b)) was obtained at a~smaller tip-sample distance, when the sharp edges in both AFM and STM channels begin to appear. More importantly the Xe tip in this experimental session was found to be positively charged \cite{Hapala_NC_2016}. 

Fig.~\ref{figPTCDAag}~(c,d) show calculated STM images obtained with the positively charged Xe tip model (Q = +0.3 elementary charge) \cite{Hapala_NC_2016}, where we considered only an $s$ orbital on the PP. The good agreement between the experimental STM images (Fig.~\ref{figPTCDAag}~(a,b)) and their theoretical counterparts (Fig.~\ref{figPTCDAag}~(c,d)) validates our approach for both far and close tip-sample distances. While in the far distance regime, the STM contrast is exclusively driven by the electronic structure of both tip and sample, in the close distance regime the distinctive sharp edges (Fig.~\ref{figPTCDAag}~(b) and (d)) coincide with the edges in the HR-AFM image \cite{Hapala_NC_2016}. The positive charge located on tip apex diminishes the apparent size of the anhydride groups at the edge of the PTCDA molecules in both STM and AFM \cite{Hapala_NC_2016}. This observation confirms that HR-STM images can potentially also be used for an~analysis of the electrostatic field \cite{Hapala_NC_2016}.

In this case, we were not able to reproduce the experimental contrast considering only a~freestanding molecule states. That is in contrast to the PTCDA/Au(111) system, where the molecular HOMO state was intrinsic to the STM signal. Here the role of the Ag substrate is very important. The simulated images were obtained at energy -0.1~eV bellow the Fermi level, where the interface states originating from hybridization of the LUMO orbital with the metallic substrate are located, see Fig.~S3~(b) in \cite{SI}. 

\begin{figure}
\centering
\includegraphics[width=8.5cm]{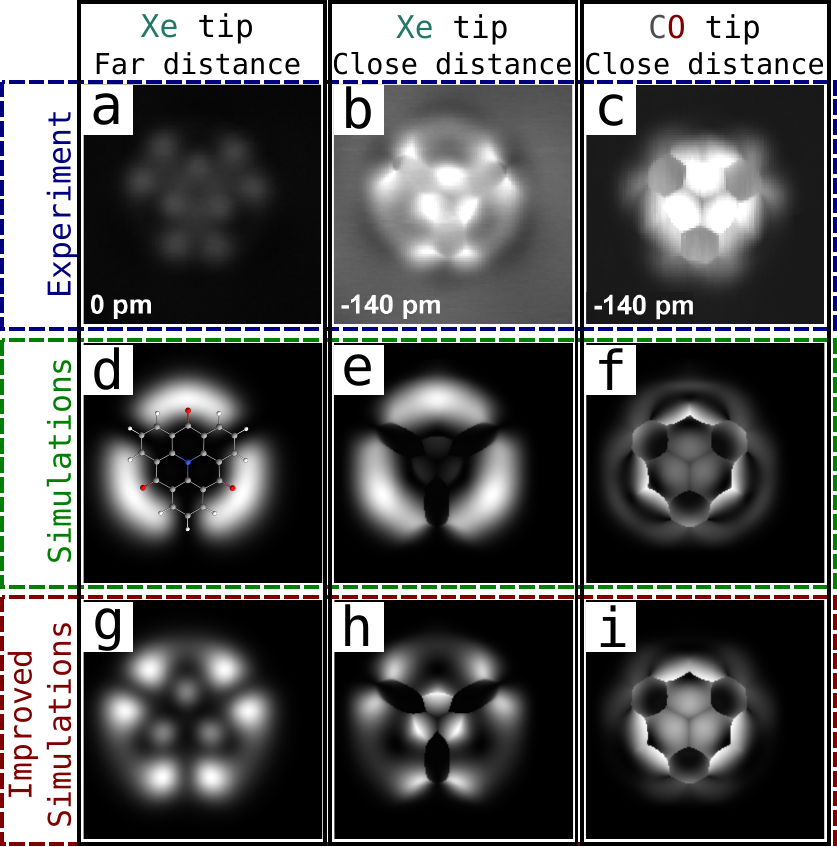}
\caption{ \textbf{Comparison of the PP-STM model with experiment - TOAT/Cu(111) - for different tips and heights:} (a-c) Experimental constant height HR-STM images of a~single TOAT molecule adsorbed on Cu(111)\cite{Nadine_submitted}: obtained with Xe tip at $V_{bias}$~=~200 mV in large tip-sample distance (a) and tip-sample distance lowered by 1.40~\AA~(b). (c) A~HR-STM image obtained with a~CO tip with small tip-sample distance at $V_{bias}$~=~100 mV. (d) and (e) PP-STM simulations at energy +0.2~eV above the Fermi level, with $s$ orbital on the PP (K~=~0.24~N/m;  Q~=~+0.3~$e$) for PP height~4.9~\AA~(d) and~3.5~\AA~(e) above the molecule. (f) PP-STM simulation at energy +0.2~eV above the Fermi level, with $p_x$ and $p_y$ orbitals on the PP (K~=~0.24~N/m;  Q~=~0.0~$e$) at height~2.9~\AA~above the molecule. (g-i) the same PP-STM simulations as (d-f),  but with $C_{a}$ constant for oxygen on the sample lowered by a factor of two. The area of all shown figures: $19~\times~19~$\AA.}
\label{figTOAT}
\end{figure}

In the last example, we will demonstrate not only that other molecules can be simulated, but also that the contrast difference between Xe and CO functionalized tips can be captured by our simulations. We will examine STM images of TOAT molecule deposited on Cu(111) surfaces, which has a~large internal charge transfer due to the presence of N atom in the center and three O atoms on the periphery of the molecule \cite{Hapala_NC_2016}.  Fig.~\ref{figTOAT}~(a,b) show STM images obtained with Xe tip in the far and close distance regimes, while Fig.~\ref{figTOAT}~(c) was acquired with a CO tip in the close distance regime \cite{Nadine_submitted}. We see that the STM contrast obtained in the close distance regime with Xe and CO tips are quite different. 

The impact of the functionalized tips is twofold. First, an additive electrostatic interaction between PP and the strong electric field of the molecule may change significantly the PP relaxation and consequently the apparent position of sharp edges \cite{Hapala_NC_2016}. In the previous work \cite{Hapala_NC_2016}, we estimated from the detailed comparison of the experimental and simulated AFM images, an~effective charge for Xe and CO to be, +0.3 and 0.0~elementary charge, respectively.  Second, the different electronic wave functions of the functional group on a~probe may change the STM contrast. In the next analysis, we will describe wave function of Xe and CO tips by $s$ and $p_x,p_y$ orbitals on PP, respectively, as they already provided the very good matches in the previous cases of PTCDA molecules.
 
Fig.~\ref{figTOAT}~(d-e) represent calculated STM images for a~positively charged Xe tip, while Fig.~\ref{figTOAT}~(f) shows a~simulated STM image using a~neutral CO tip, both at energy +0.2~eV above the Femi level. The effect of different effective charge can be nicely seen from a~different apparent shape of outer benzene rings in the close distance images (Fig.~\ref{figTOAT}~(b,c)). Different orbital symmetries of the tip wave function give rise to distinct contrast in the STM images (compare e.g. different contrast in the central part and on periphery of the molecule). However, overall agreement between the experiment and the simulation is not very good, especially on the periphery of the molecule.

One possible explanation can be related to the more complex electronic structure of the probe. In particular, Gross et al. \cite{Gross2011} achieved good agreement between STM images acquired with CO tips in far distance regime by taking into account linear combination of $s$, $p_x$ and $p_y$ orbitals on the probe. Pavl{\'i}{\v{c}}ek et al. \cite{Pavlicek_PRL_2013} claimed that the $p$ and $s$ contributions can depend on the applied bias voltage. However, in our case, the combination of $s-p$ orbitals does not improve the agreement with the experimental data.   

On the other hand, from analysis of the Hartree potential above the adsorbed molecule, we found out a~local increase of the tunneling barrier above the oxygens due to a~negative partial charge on the oxygen (see Fig.~S4~(c)). We tried to mimic the variation of the tunneling barrier by lowering the $C_a$ constant for the oxygen atoms by a factor of two. The HR-STM simulations displayed in Fig.~\ref{figTOAT}~(g-i)  show significant improvements in the match with the experimental figures for both tip functionalizations and for both tip-sample distance regimes. 

More rigorous treatment of the $C_a$ constant, by taking into consideration the local variation of the potential barrier height, is left as a subject for further development.
We would like to emphasize, that many other parameters (e.g. chemical and atomic structure of the whole tip apex including metallic base, charges induced due to the applied bias voltage and/or tip-sample proximity \cite{Corso_PRL2015}, and tunneling between the PP and the tip \cite{Hapala_PRB_2014}), which are not taken into account in our model, could also play role in varying the STM contrast. Despite this fact, we found the PP-STM model provides satisfactory agreement with the available experimental data.

In conclusion, we have introduced the PP-STM model for simulations of HR-STM images acquired with a flexible tip apex. The PP-STM model takes into account both the relaxation of the probe particle and the tunneling process between electronic states of the sample and the tip. We have employed the Bardeen theory to describe the tunneling process, while the relaxation of the probe particle is described by the mechanistic PP-AFM model \cite{Hapala_PRB_2014, Hapala_PRL_2014}. We have performed extensive comparison of simulated HR-STM images with experimental evidence to demonstrate the validity and the limits of the PP-STM model. The model sheds more light into HR-STM mechanism, which consists of the standard STM imaging heavily distorted by the relaxation of the flexible tip apex. We believe that the detailed understanding of the  high-resolution mechanism of STM imaging will serve to further proliferation of wider application of this technique.         

\section{ acknowledgements }

We would like to thank to R. Temirov (PTCDA/Au(111)); O. Stetsovych and M. {\v{S}}vec (PTCDA/Ag(111));  J. van der Lit, N. J. van der Heijden and I. Swart (TOAT/Cu(111)) for providing us the experimental images and discussion. We would like to thank to N. Pavlicek and J. Repp for fruitful discussions. We acknowledge the support by GA\v{C}R, grant no.\ 14-16963J. 

%

\end{document}